# Fluctuation Assisted Ejection of DNA From Bacteriophage

Michael J. Harrison

*Department of Physics and Astronomy, Michigan State University*
*East Lansing, MI 48824-2320, USA*

**Abstract.** The role of thermal pressure fluctuations in the ejection of tightly packaged DNA from protein capsid shells is discussed in a model calculation. At equilibrium before ejection we assume the DNA is folded many times into a bundle of parallel segments that forms an equilibrium conformation at minimum free energy, which presses tightly against internal capsid walls. Using a canonical ensemble at temperature T we calculate internal pressure fluctuations against a slowly moving or static capsid mantle for an elastic continuum model of the folded DNA bundle. It is found that fluctuating pressure on the capsid mantle from thermal excitation of longitudinal acoustic vibrations in the bundle may have root-mean-square values which are several tens of atmospheres for typically small phage dimensions.

## INTRODUCTION

It has been emphasized that thermal fluctuations away from equilibrium in small systems assume increasingly important roles as the system size becomes ever smaller and approaches molecular dimensions [1]. In the case of viral entities, where the phage size may very considerably [2,3], thermal pressure fluctuations may become exerted internally by confined DNA that is already tightly packged in an equilibrium conformation within a capsid mantle. Such pressure fluctuations can be expected to play a signigicant role in the initial packaging and subsequent dynamics of DNA ejection from bacteriophage [4,5,6,7,8] when the virus engages a host cell surface receptor and transfers its DNA into the cell.

Geometrical and topological constraints on the structure and size of viral capsids were discussed in an early paper by Caspar and Klug [9] almost fifty years ago. More recent work [2,10,11] has discussed highly symmetric capsid structure in the form of nanometer-sized protein shells. And viral infectivity has been related to DNA length and capsid size [12]. It has also been concluded that very large conformal changes may occur in some capsid shells when their viral genomes become tightly packaged [13]. We shall adopt a simpler model of viral capsids that contain packaged genomes than has generally been discussed [14].

## CALCULATION OF MODEL ENCAPSIDATED DNA AND CAPSID

Consider a cylindrical bundle of folded double-stranded DNA bacteriophage segments with circular cross-section diameter L and equal segment lengths L, packaged coaxially and situated within a cylindrical capsid cavity of diameter (L+2h) and length L. The bundle of DNA segments is then encased by an annular cylindrical shell of thickness h. If the cavity is to have twice the volume [15] of the DNA bundle, then geometric considerations lead to the relation $h = L / [2(1+\sqrt{2})] = 0.2071/L$, so that $(L+2h) = 1.414\ L$ is the capsid diameter. Capsids are nanometer sized in general magnitude [10], but vary greatly [2,11] depending specifically on the DNA genomes they carry within. We shall adopt a representative average capsid diameter of 48 nm [5,21] in our calculation of fluctuation pressures exerted between a vibrating DNA bundle and the capsid shell to which it is tethered.

Our discussion of the role of fluctuations rests on a picture of a tightly packaged encapsidated DNA bundle that acquires its thermal equilibrium structure through variational minimization of the free energy [7] of the capsid-DNA system. Fluctuations of the DNA bundle about its equilibrium conformation takes place in the form of thermally excited longitudinal acoustic vibrations. These vibratory DNA excitations will be regarded as taking place bounded by capsids which themselves continue to maintain configurations close to their equilibrium geometry, and do not participate in thermal excitations on the same more rapid time scales as the DNA bundles. The bundles of DNA segments are assumed to be connected internally at both ends to their capsid mantles.

In order to calculate the thermal fluctuations in pressure at the locations where DNA bundles are attached to their capsid shells we shall approximate a bundle of folded DNA segments by a continuum elastic rod [16, 17]. Longitudinal acoustic vibrations of such a rod fixed at both ends to a static capsid inner surface depend on the longitudinal sound velocity within the rod, which in turn depends on the rod's macroscopic mass density and elastic properties. These two quantities enter the speed of longitudinal sound according to $v = \sqrt{(Y/\rho)}$ , where Y is the bulk modulus and $\rho$ is the mass density of the DNA bundle.

We relate these quantities to the microscopic force constant and total genome mass of single double -stranded DNA molecules which have become folded into bundles of length L and circular cross-section area $\pi(L/2)^2$ . The elastic response of individual double-stranded DNA molecules has been measured [18]. And elastic constants have been used to calculate phase velocities of sound waves [19]. Brillouin scattering has been used to determine the longitudinal velocity of sound in B-DNA fibers over a quarter century ago [20] with the result v = 1.9 km/s, which we shall adopt for the present calculation. The longitudinal sound velocity in a bundle of fibers regarded as a continuum coincides with the sound velocity of a single fiber since Y and $\rho$ have the same ratio.

We introduce a coordinate system with its origin on the cylindrical bundle's axis where it connects with a static capsid mantle. The x-axis coincides with the bundle's axis and extends through it to the other end of the tightly packaged bundle on the static capsid's internal surface a distance L away. Then with y(x,t) representing a general longitudinal displacement field, for t > 0 in a continuum bundle we have

$$y(x,t) = \sum_n \alpha_n(t) \sin(n\pi x / 2L) \quad , \qquad 1.$$

which obeys boundary conditions $y(0,t) = 0$ and $y(L,t) = 0$ , where n are even integers and $\alpha_n(t)$ are normal coordinates for longitudinal wave motion in the folded DNA bundle, regarded as a continuum. We take

y(x,t) = 0 for t <0. The total hamiltonian H of the wave system is given by the sum of its kinetic and potential energy:

$$H = \int_0^L dx \ [s\rho/2] \, [(\partial y/\partial t)^2 + v^2 (\partial y/\partial x)^2 ] \quad , \qquad 2.$$

where $s = \pi(L/2)^2$ is the cross-sectional area of the packaged DNA bundle, ρ is the mass density and v is the longitudinal sound velocity.

We substitute Eq.(1) into Eq.(2) and obtain

$$H = (s\rho L/4) \sum_n |\alpha_n'|^2 + ( v^2\rho s\pi^2/16L) \sum_n |\alpha_n|^2 n^2 \ . \qquad 3.$$

The total energy H depends only quadratically on the $\alpha_n(t)$ and their time derivatives. If we now adopt a canonical ensemble to obtain the thermal average <H> at temperature T, each quadratic term in H has its equipartition average value kT/2 for the ensemble. We obtain:

$$< |\alpha_n|^2> = [8LkT/ (n^2v^2\rho s\pi^2)] \ , \qquad 4.$$

where the brackets denote the thermal average.

At the closed end pressure antinode, x = 0, the pressure fluctuation against the constraining internal wall of the capsid is

$$\Delta p(0,t) = - v^2 \rho \, (\partial y/\partial x \,|_{x=0}) \qquad 5.$$

for a displacement field y(x,t). We now take the ensemble average $< |\Delta p|^2 >$ and obtain

$$< |\Delta p|^2 > \ = \ [v^4\rho^2 \pi^2 /(4L^2)] \sum_{nm} nm < \alpha_n\alpha_m > \ . \qquad 6.$$

But $< \alpha_n\alpha_m > = <| \alpha_n |^2> \delta_{nm}$ since in thermal equilibrium the normal coordinates $\alpha_n$ are uncorrelated with respect to their time dependence. A single sum results:

$$< |\Delta p|^2 > \ = \ [ v^4\rho^2 \pi^2 /(4L^2)] \sum_n n^2 < |\alpha_n|^2 > \ . \qquad 7.$$

Substituting Eq.(4) into Eq.(7) we obtain

$$< |\Delta p|^2 > \ = [2v^2\rho kT/(Ls)] \sum_n 1 \ . \qquad 8.$$

There must be a cut-off limit in Eq.(8) reflecting a short wavelength limit on well-defined longitudinal sound waves in the DNA packaged bundle, which corresponds to a highest even mode number N in the Eq,(8) summation:

$$\sum_n^N 1 = \ N/2 \ . \qquad 9.$$

The thermal fluctuation noise pressure on the capsid mantle is then given by:

$$< |\Delta p|^2 > = [N v^2 \rho kT / (Ls)] . \qquad 10.$$

We obtain a *lower bound* to the fluctuating pressure magnitude for N=2. Taking M=ρ(Ls) the total genome mass within the capsid, we define the root-mean-square fluctuating pressure magnitudes $P_{rms} \equiv \sqrt{(<|\Delta p|^2>)}$ and obtain

$$P_{rms} \geq \sqrt{[2v^2MkT/(Ls)^2]} \quad . \qquad 11.$$

Since M is proportional to the genome length, Eq.(11) suggests that the fluctuating pressure magnitude at temperature T should be proportional to the square root of the genome length divided by the capsid volume whenever the latter is a multiple of the packaged genome volume.

For our model genome package $s = \pi(L/2)^2$ and the folded cylindrical bundle has half the volume of the coaxial caspid cavity. We then have $(Ls) = 0.785\ L^3$ and $(L+2h) = L[1+1/(1+\sqrt{2})]$ as the capsid diameter, which we take as 48 nm [21]. Solving for L we obtain L = 33.94 nm, so that the bundle volume is $(Ls) = 3.07 \times 10^{-17}\ cm^3$. Using the virus ϕ29 as an example [21] we take the genome length to be 6.6 – μm . With a base pair separation of 3.4 Å we have the number of base pairs in the genome equal to $1.94 \times 10^4$ for ϕ29. Multiplying the number of base pairs by the average mass of a base pair, $1.096 \times 10^{-21}$ grams, we obtain the genome mass M = $2.13 \times 10^{-17}$ grams. For T = 300 Kelvin and v = 1.9 km/s Eq.(11) then gives $P_{rms} \geq 8.11$ atm. A virus genome that is sixty times longer, such as Mimivirus, would develop an rms pressure of $P_{rms} \geq 63.3$ atm. The dependence of ejection forces on genome length has been noted in bacteriophage lambda [22]

In consequence of the above calculations we conjecture that time-dependent fluctuation pressures leading to $P_{rms}$ act like a trigger that assists the internal forces exerted on tightly packaged DNA to eject the genome.

I wish to thank Professor Lisa Lapidus for several stimulating conversations.

**REFERENCES**

1. C. Bustamente, J. Liphardt, and F. Ritort, Physics Today **58 (**7**),** July 2005, p.43.
2. R.V. Mannige and C. L. Brooks III, PNAS **106 (**21), May 26, 2009, p.8531.
3. R. Mannige and C. Brooks III, Phys. Rev. E **77**, 051902.
4. M. M. Inamdar, W.M. Gelbart, and R. Phillips, Biophys. J. **91**, July 2006, p.411.
5. P. K.Purohit, J. Kondev, and R. Phillips, PNAS **100 (6),** March 18, 2003, p.3173.
6. P. Grayson, A. Evilevitch, M. M. Inamdar, P. K. Purohit, W. M. Gelbart, C. M. Knobler, and R. Phillips, Virology **348**, (2006), p.430
7. S. Tzlil, J. T. Kindt, W. M. Gelbart, and A. Ben-Shaul, Biophys. J. **84** (3), March 2003, p.1616.

8. L. Ponchon, S, Mangenot, P. Boulanger, and L. Letellier, Biochim. Biophys. Acta **1724** (3), Aug. 5, 2005, p. 255.
9. D. L. D. Caspar and A. Klug, Cold Spring Harbor Symposium on Quantitative Biology **27, (**1962), p.1
10. M. M. Gibbons and W. S. Klug, J.Mater.Sci **42**,(2007), p.8995.
11. O. M. Elrad and M. F. Hagan, Nano. Lett. **8** (11), (2008), p.3850.
12. A. Evilevitch, Q. Rev. Biophys. **40**, (2007), Cambridge Univ. Press, p.327.
13. I.Gertsman, L.Gan, M.Guttman, K.Lee, J.A Speir, R.L.Duda, R.W.Hendrix, E.A.Komives, and J.E.Johnson, Nature **458**, 2 April 2009, p.646.
14. W. Lucas and D.M. Knipe, Encycl. Life Sci., 2002, Macmillan Pub. Ltd.
15. M.Sun and P.Serwer, Biophys. J. **72**, February 1997, p.958.
16. B. Eslami-Mossallam and M.R. Ejtehadi, Phys. Rev.E **80**, (2009), p. 011919.
17. R.S. Manning, J.H. Maddocks, and J.D. Kahn, J. Chem. Phys.**105** (1996), p.5626.
18. S.B. Smith, Y. Cui, and C. Bustamante, Science **271**, 9 February 1996, p. 795.
19. T.C. Bishop and O.O. Zhmudsky, arXiv: physics/0108008v1 [physics. bio-ph] 6 August 2001.
20. M.B. Hakim, S.M. Lindsay, and J. Powell, Biopolymers **23**, (1984), p.1185.
21. D.E. Smith, S.J. Tans, S.B. Smith, S. Grimes, D.L. Anderson, and C. Bustamante, Nature **413**, 18 October 2001, p.748.
22. P. Grayson, A, Evilevitch, M.M. Inamdar, P.K. Purohit, W.M. Gelbart, C.M. Knobler, and R. Phillips, Virology **348**(2), 10 May 2006, p.430.